\documentclass[letterpaper,twocolumn,10pt]{article}
\usepackage{Style/usenix2021_SOUPS}

\usepackage{tikz}
\usepackage{amsmath}
\usepackage{stfloats}
\usepackage{caption}
\usepackage{subcaption}
\usepackage{tcolorbox}
\usepackage{graphics}      
\usepackage[T1]{fontenc}   
\usepackage{mathptmx}
\usepackage{booktabs}
\usepackage{textcomp}
\usepackage{xspace}

\usepackage{microtype}        
\usepackage[all]{hypcap}    
\usepackage{ccicons}          
\usepackage{todonotes}
\makeatletter
\newcommand*{\etal}{{\em et al.}\@\xspace}
\newcommand*{\eg}{{\em e.g.,}\@\xspace}
\newcommand*{\ie}{{\em i.e.,}\@\xspace}
\newcommand{\obf}{{\fontfamily{cmss}\selectfont{Obfuscator}}\xspace}
\newcommand{\bout}{{\fontfamily{cmss}\selectfont{PowerCut}}\xspace}
\newcommand{\va}{smart speaker\xspace}
\newcommand{\vas}{smart speakers\xspace}
\newcommand{\vaT}{Smart speakers\xspace}
\newcommand{\varun}[1]{\textcolor{black}{#1}}

\newcommand{\obfT}{{\fontfamily{cmss}\selectfont{Obfuscator}}\xspace}
\newcommand{\boutT}{{\fontfamily{cmss}\selectfont{PowerCut}}\xspace}
\usepackage{pifont}
\newcommand{\cmark}{\ding{51}}%
\newcommand{\xmark}{\ding{55}}%

\begin{document}

\date{}

\title{PowerCut and Obfuscator: An Exploration of the Design Space for \\ Privacy-Preserving Interventions for Smart Speakers}

\def\plainauthor{Varun Chandrasekaran, Suman Banerjee, Bilge Mutlu, Kassem Fawaz}

\author{Varun Chandrasekaran, Suman Banerjee, Bilge Mutlu, Kassem Fawaz \\ University of Wisconsin-Madison} 


\maketitle
\thecopyright

\begin{abstract}
{
The pervasive use of \vas has raised numerous privacy concerns. While work to date provides an understanding of user perceptions of these threats, limited research focuses on how we can mitigate these concerns, either through redesigning the \va or through dedicated privacy-preserving interventions. In this paper, we present the design and prototyping of two privacy-preserving interventions: `\obfT' targeted at disabling recording at the microphones, and `\boutT' targeted at disabling power to the \va. We present our findings from a technology probe study involving 24 households that interacted with our prototypes; the primary objective was to gain a better understanding of the design space for technological interventions that might address these concerns. Our data and findings reveal complex trade-offs among utility, privacy, and usability and stresses the importance of multi-functionality, aesthetics, ease-of-use, and form factor. We discuss the implications of our findings for the development of subsequent interventions and the future design of \vas.
}
\end{abstract}

\section{Introduction}

\begin{figure}[t]
\centering
  \includegraphics[width=\columnwidth]{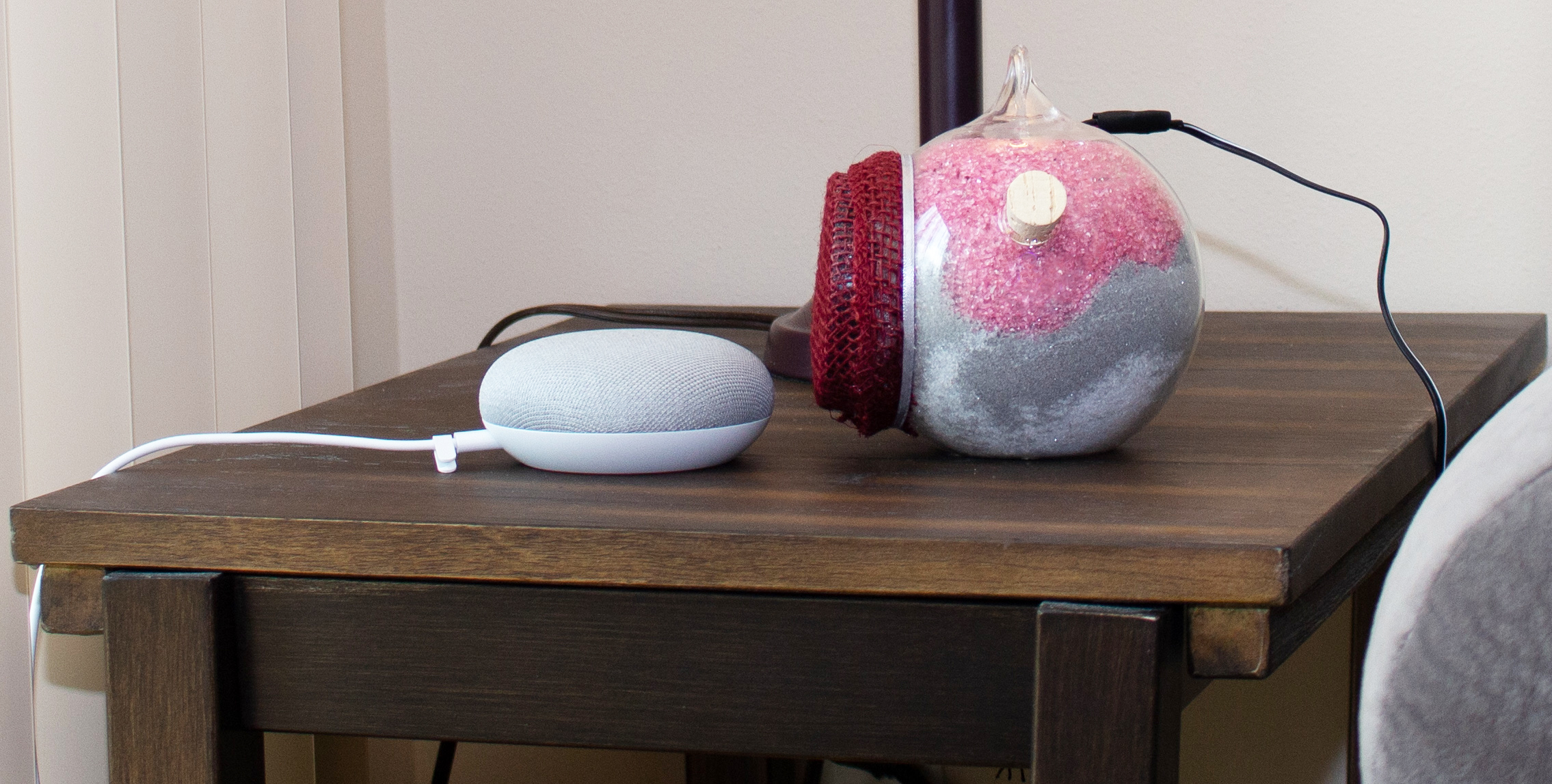}
  \caption{The \obf design probe next to a Google Home Mini Device. \obf uses ultrasound jamming to prevent the \va from listening to the user's conversations and is designed to appear as a tabletop ``trinket'' to blend into the user's home environment.}
  \label{fig:teaser}
\end{figure} 

\vaT, or network-connected speakers with integrated virtual assistants, are becoming increasingly pervasive in households. In 2020, nearly 90 million US adults used a \vaT~\cite{adult}. \vaT offer their users a convenient way to access information, set alarms, play games, or set to-do lists. \vaT also integrate with other devices to realize smart home applications. However, this convenience comes at a potential privacy cost; these devices operate in an always-on mode at earshot of nearby conversations.

Smart speakers already provision built-in privacy controls; they are supposed to process audio inputs locally until they detect a wake word, and they pack a button that mutes their internal microphone. Unfortunately, both provisions are not very effective at protecting the user's privacy. Recent incidents raise concerns about {\em passive} privacy threats~\cite{abdi2019more,laualexa,malkin2019privacy,jin2018they}. \vaT can be mistakenly triggered without the presence of a wake word ~\cite{alexa_recording, kumar2018skill,dubois2020speakers}, causing it to record speech not intended as commands. Further, security researchers have documented {\em active} vulnerabilities that indicate the potential for malicious exploitation of  \vas~\cite{alexa_compromise,gizmodo,google, apple,zhang2017dolphinattack}. Further, the effectiveness of the mute button to address these problems is in doubt~\cite{laualexa}. Recent studies, including the one in this work, indicate that users find this button inconvenient to utilize and not trustworthy in some cases~\cite{sensys_jamming}. While different technical interventions have been proposed recently~\cite{sensys_jamming, chenwearable}, the design space for such interventions remains under-explored. This paper contributes to an improved understanding of the design elements and understanding user experience with these interventions. 

In our work, we aim to understand better the user perceptions around the potential technological solutions to the privacy issues involving \vas through a technology probe-based approach~\cite{Hutchinson:2003}. The objective of our study is not to validate particular design choices but to understand user perceptions of such interventions better and extract design requirements for them. We utilize the \va's {\em built-in} mute button as a baseline, to understand user perceptions of how device manufacturers provide privacy control. We utilized two technology probes to represent  {\em bolt-on} privacy-preserving interventions: (a) \bout, a smart plug that allows the user to engage/disengage the power supply to a \va remotely, and (b) \obf (Figure~\ref{fig:teaser}), which uses ultrasound to deafen the \va's microphone, preventing the \va from listening to nearby conversations. The probes intercept two key resources required for successful \va functionality: {\em power} (for basic operation) and {\em microphone inputs} (for voice-based interaction).

To promote user reflection on our privacy-preserving interventions, we conducted in-home demonstrations of our technology probes through in-depth interviews at 24 households. Our interviews took place over two phases within a year (July 2018 -- August 2019), providing us with insight into how such perceptions and attitudes might change over time. Our interviews involved users with diverse demographics, including {\em casual} (or recreational) users and {\em power} (or proficient) users, enabling us to distinguish perceptions and design requirements for different user groups. Our findings highlight a complex trade-off between privacy, utility, and usability: the interventions (a) should be plug-and-playable \ie require minimal setup and upkeep, (b) have a small physical footprint and fit within its environment, (c) offer additional features beyond privacy preservation, (d) does not affect the interaction model with the \va, and (e) must survive the test of time \ie it should be compatible with existing and future iterations of \vas. Through this work, we present the design of our technology probes and our in-home study, and discuss our findings. We conclude with a discussion of their implications for the \vas as well as other privacy-sensitive technologies.

\section{Background}
\label{sec:background}
\textcolor{black}{
Our study considers \vas deployed in home environments, focusing on (a) Google Home Mini and (b) Amazon Echo Dot as described in Table~\ref{table:features}. Users interact with these devices to achieve a multitude of tasks, such as information access, interaction with other smart devices, setting alarms/timers, and voice calls. A typical interaction with a \va starts with the user speaking a wake word, such as ``{\em ...Alexa}'' or ``{\em ...Google.}'' Upon recognition of the wake word, the device indicates its readiness to receive command through a visual cue. Then, the device sends the speech segment to the cloud, which verifies the wake word and processes the accompanying command~\cite{schonherr2020unacceptable}. Verification is necessary since on-device models are typically less accurate to minimize their compute footprint and latency of predictions~\cite{apple_2step, google2step}. 
} 
\textcolor{black}{ As such, the \va has to be always on, continuously listening for a user to speak the wake word. Ideally, the device should only record, and communicate to its cloud, the commands that were triggered by a wake word. In many circumstances, however, the device's operation might not match its expected behavior. This results in the two privacy threats described below.} Note, these threats also provide context about scenarios where we envision privacy-preserving interventions to be used.

\begin{table}[h!]
\small
\begin{center}
  \begin{tabular}{  p{2.6cm}  p{2.6cm}  p{2.1cm}}
    \toprule
    \textbf{Feature} & \textbf{Home Mini} & \textbf{Echo Dot} \\
    \midrule
    Manufacturer & Google & Amazon \\
    Height $\times$ Diameter & 4.3 $\times$ 9.9 cm & 3.3 $\times$ 7.6 cm  \\
    Wake words & "... Google" & "... Alexa" \\
    Visual Cue & Dots on the surface & LED band \\
\textcolor{black}{ Privacy controls} & Mute button & On/Off switch  \\
    \bottomrule
    \end{tabular}
\caption{Salient features of \vas in 2019. \vspace*{-4mm}}
\label{table:features}
\end{center}
\end{table}

\noindent{\bf \varun{Passive Threats:}} The first threat occurs due to innocuous and inadvertent recording \ie when the \va misunderstands ongoing conversations to contain the wake word. Recent analysis~\cite{dubois2020speakers} reported that everyday phrases, such as those from TV shows, can accidentally activate a \va, resulting in 10 seconds of speech being sent to the cloud. There have been several incidents where these devices have exported user conversation, including those not preceded with a wake word. While one organization claims this is a one-off act \cite{alexa_recording}, another blames erroneous code~\cite{google, google-bug}. There have also been reported instances where several organizations hired human contractors to listen and tag different recordings from these devices, which include commands and non-commands~\cite{apple}; this is a severe deviation from perceived device operation. Collectively, we refer to these violations as {\em passive} privacy threats.

\noindent{\bf \varun{Active Threats:}} The second occurs due to compromise of the actual device or its operation. A malicious entity can compromise the software running on the connected \va to turn it into a listening device. Such an entity can also change the operation of the device through developing applications that record the user's conversations~\cite{zhang2019dangerous,kumar2018skill,alexa_compromise} or inject stealthy commands to wake up the device without the user's awareness~\cite{roy2018inaudible,zhang2017dolphinattack,carlini2016hidden}. Since these devices are connected to the internet, such alterations are capable of extracting various forms of sensitive information. We refer to such threats as {\em active} privacy threats.




\section{Methodology}
\label{methodology}

We envision privacy-preserving interventions to address potential passive and active threats, especially in scenarios that users perceive as sensitive. Such scenarios can include users receiving visitors or having sensitive conversations. Concretely, there exist two strategies to safeguard users' privacy in sensitive scenarios: (a) redesigning the \va to provide provable privacy guarantees, or (b) designing interventions that co-exist with the \va. The former is a challenging proposition as most of the software and hardware required for successful \va functioning is proprietary. Additionally, it would involve trusting the device provider (a theme that will revisit later) to provide proof that the user's privacy was violated. 

To this end, we explore the design of {\em bolt-on, hardware-based} interventions. These \varun{interventions} are less abstract than software-based ones; they allow the users to physically and directly interact with them. For thoroughness, we compare and contrast our findings with the usage of a {\em built-in} feature found in \vas --- the mute button. The results of our research inform the design of \vas with improved privacy properties and privacy-preserving interventions in physical spaces. Note that our analysis is restricted to \vas and not smartphones (which are also susceptible to the threats discussed earlier). In particular, \vas are {\em easier to protect} as they are less {\em mobile} than smart-phones.

\noindent{\bf What is a tech probe?} We follow a \textit{technology probe}-based design approach, which allows us to identify design guidelines that capture the users' mental models. We aim to understand how the users of \vas react to different privacy-enhancing technologies using proof-of-concept prototypes (or probes). In a technology probe, the researcher develops an interface that packages the core functionality of the privacy intervention. The researcher keeps the interface as simple as possible to avoid making design decisions~\cite{buchenau2000experience,odom:2012}. When an individual interacts with this basic interface, the researcher {\em probes} the individual to reveal a specific phenomenon that is otherwise hidden~\cite{Hutchinson:2003}. 

{\em In our case, we probe and interview the users to elicit their immediate reactions and reflections about what design elements are missing and need to be introduced}. We follow with qualitative analysis to reveal the design guidelines for a privacy intervention in the \va environment. In follow-up work, we are planning to realize the privacy intervention and set up a diary study to understand longer-term use. This will allow us to concretely measure any issues users have with the actual intervention that was conceptualized for deployment. A note on the nomenclature: in this work, we design technology probes (or probes for short) to elicit insight about the final intervention (which we do not design), but for which we make recommendations.


\subsection{Iterative Design Process}

In designing our technology probes, we followed an iterative design process. We first explored the broad space of solutions (presented in Table \ref{table:strawman}), their efficacy against an adaptive adversary, and discussed the advantages and disadvantages of each approach. Recall that our objectives are to design an easy-to-use intervention with intuitive yet provable privacy guarantees. It is clear that modifying device hardware and controlling network flow does not provide the desired privacy protection -- the encrypted nature of network traffic makes it difficult to tag and discard packets (with information) that are not to be shared, while inadvertent \va activation will persist. One could change the wake word to reduce the frequency of spurious activation/recording. However, this phenomenon is not well understood for it to be a definitive fix, and a harder-to-pronounce wake word has usability problems. 

\begin{table}[h!]
\small
\begin{center}
  \begin{tabular}{  l c c }
    \toprule
    \textbf{Possible Solutions} & \textbf{{\em Active} Threats} & \textbf{ {\em Passive} Threats} \\
    \midrule
Network interception  & \xmark & \xmark \\ 
Hardware modifications & \xmark & \xmark \\ 
Change the wake word & \xmark & \cmark \\ 
Discard \va & \cmark & \cmark \\
    \bottomrule
    \end{tabular}
\caption{Space of possible solutions and their effectiveness against malicious programming (or {\em active} privacy threats) and inadvertent recording (or {\em passive} privacy threats).}
\label{table:strawman}
\end{center}
\end{table}

Observe that while some of our possible solutions are intuitive to the average user, others (such as network monitoring) are not.  Based on preliminary discussion with several end-users, we converged on a set of dimensions that we found relevant to the final design of our probes. They are (a) the method of user-probe interaction \ie hands-free vs. physical, (b) the ease of deployment, and (c) the ease of understanding the privacy properties the probe provides. We stress that these dimensions are not exhaustive and merely serve as a starting point for our design.

We construct two probes guided by these suggestions. Again, we stress that we do not seek to evaluate the efficacy of these probes in preserving privacy. We do not attempt to understand how people use these probes as well. Doing so requires running a diary study with the probe deployed in users' homes.  We describe the probes used in our study, including those we conceptualized, below. We also briefly state our analysis of the trade-offs ensuing from each probe.

\begin{figure}[!b]%
\centering
    \begin{subfigure}{0.25\columnwidth}
        \centering
        \includegraphics[width=\textwidth]{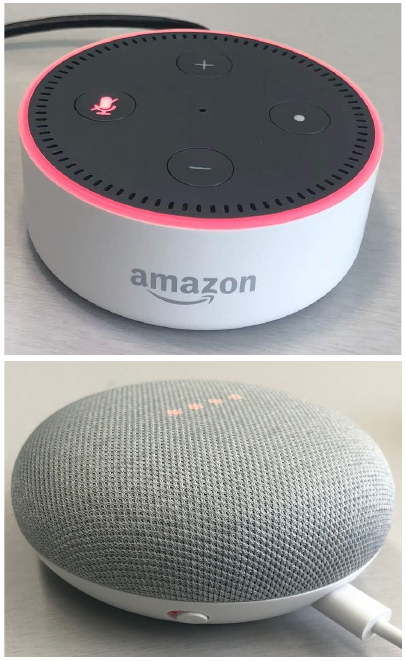}
        \caption{Mute}
        \label{fig:mute}%
    \end{subfigure}
    \begin{subfigure}{0.3\columnwidth}
        \centering
        \includegraphics[width=\textwidth]{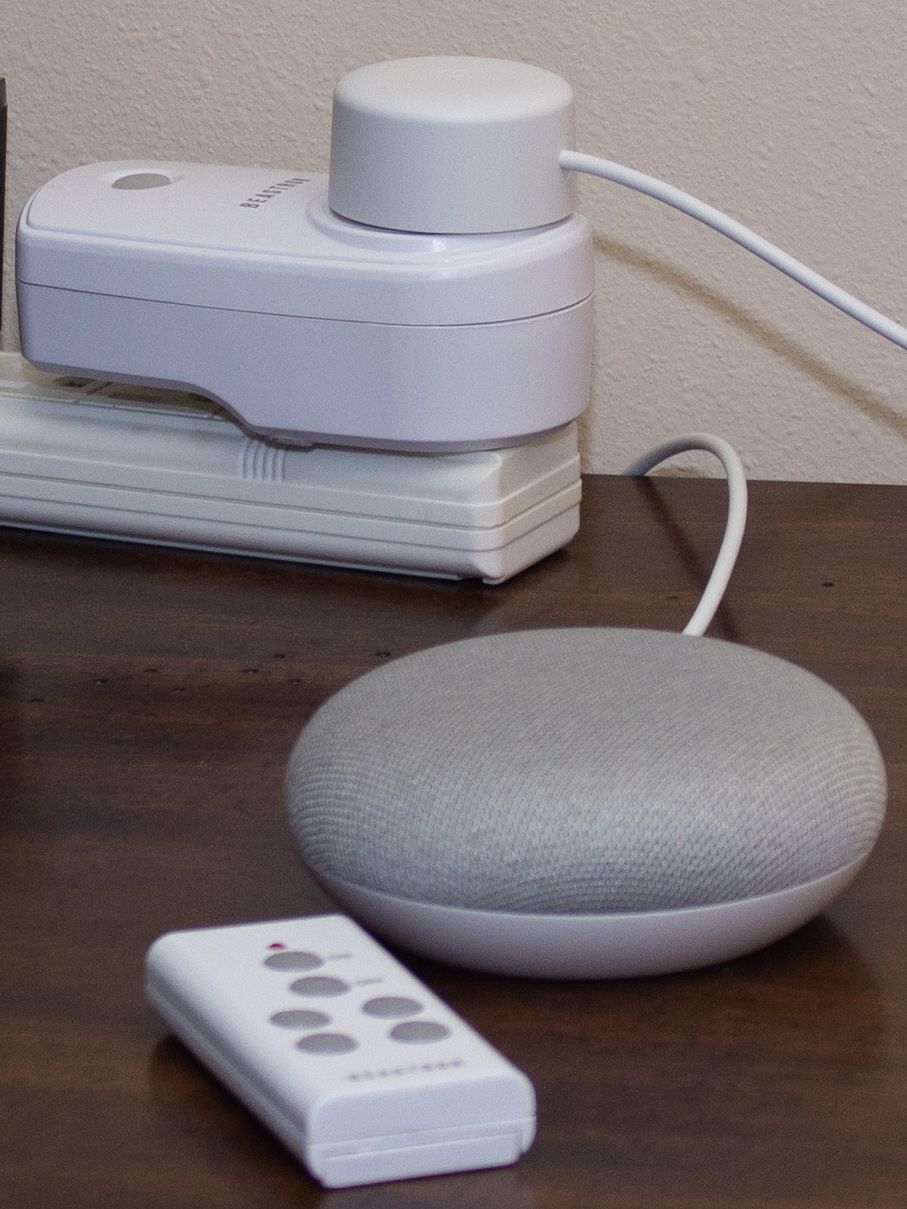}
        \caption{\bout}
        \label{fig:plug}%
    \end{subfigure}
    \begin{subfigure}{0.4\columnwidth}
        \centering
        \includegraphics[width=\textwidth]{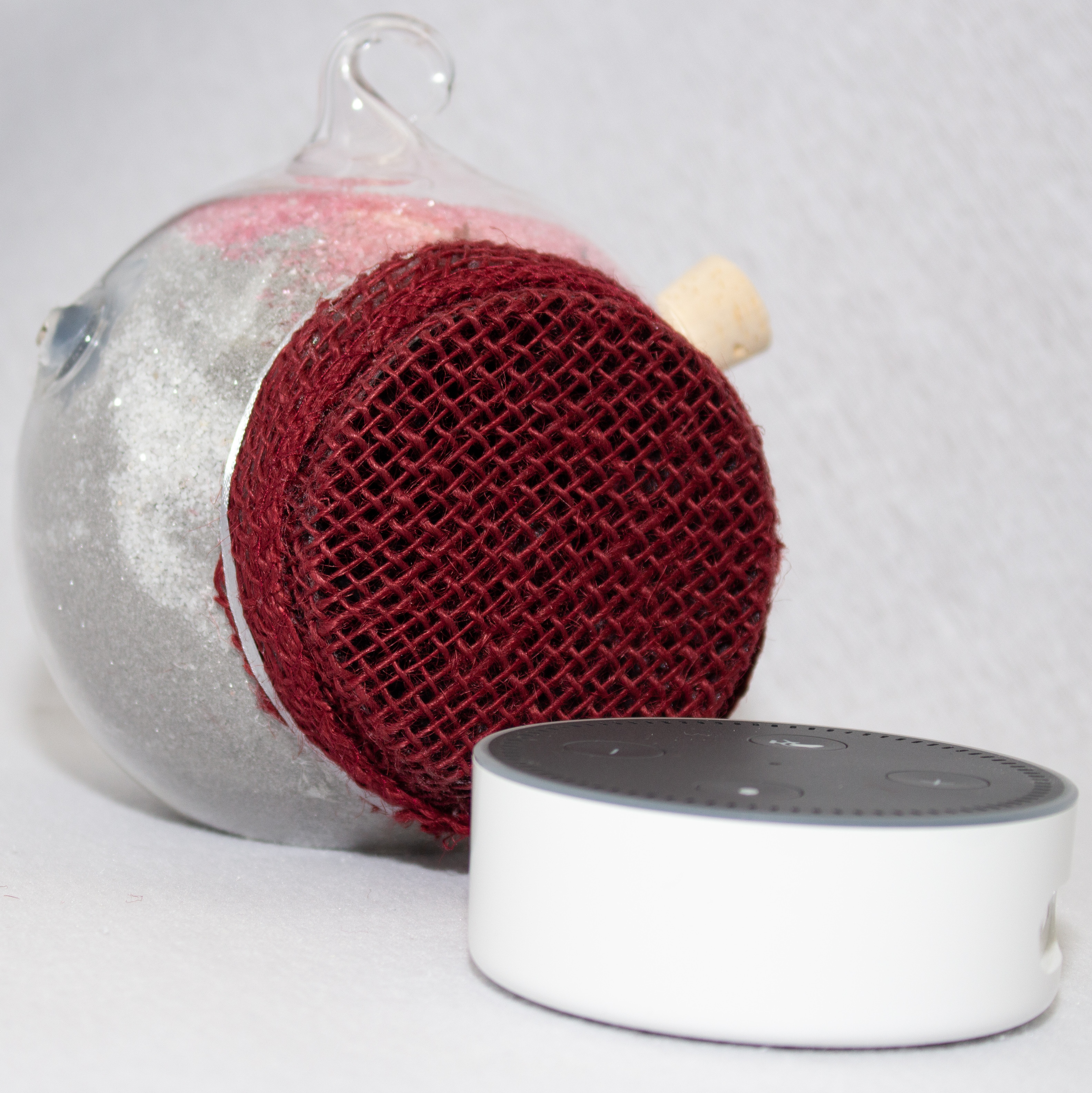}
        \caption{\obf}
        \label{fig:jam}%
    \end{subfigure}
    \caption{The three employed privacy probes.}
    \label{fig:all_interventions}
\end{figure}

\vspace{1mm}
\noindent{\textbf{1. Mute:}} The ``mute'' feature represents a \textit{built-in} privacy control (Figure~\ref{fig:mute}). It is available as a push button on the top panel of some of the Amazon Echo Dots and as a sliding button on the side of some of the Google Home Minis. The device manufacturers state that the microphone is deactivated when the mute button is turned on ({\em c.f.} Figure~\ref{fig:mute}). Naturally, activating the mute button stops the \va from responding to the user's voice commands. Upon activation, the Echo Dot's ring color changes to red, and the four lights atop the Google Home Mini turn red. 

\noindent{\em Trade-offs:} While inbuilt, the mute feature requires the user to physically interact with the device to engage the control. It also requires the user to place trust in the manufacturer's implementation of the feature.

\vspace{1mm}
\noindent{\textbf{2. {\boutT}:}} While the mute button focuses on disengaging the microphone inputs, we conceptualize another probe to disengage the electricity supply. A naive way of achieving our goal is to either disconnect the \va's cord from the outlet or disconnect the cord connected to the \va. However, both options involve physical interaction with the device. Thus, we use a remote-controlled outlet\footnote{Beastron Remote Controlled Outlet}(Figure~\ref{fig:plug}). The user deploys \bout by connecting the \va to the outlet through the smart plug (as seen in Figure \ref{fig:plug}). 

\noindent{\em Trade-offs:} We use a commercial smart-plug because we believe that users will be familiar with such products, minimizing their time for acclimatization. Additionally, we speculate that users will trust the functionality of such widely-used products, with no negative publicity. \boutT is conspicuous and rugged; we believe that its form factor makes it easier to understand and use. The user can engage/disengage \bout through a remote control (with a range of operation of 100 feet) without the need to physically interact with the device. Additionally, the smart plug we chose provides a visual cue --- an LED glows {\em red} when powered on to indicate that the \va is active.  Clearly, \bout offers immediate privacy guarantees. This comes at a cost; the users have to wait for a lengthy boot time whenever they wish to reuse the \va. Additionally, the form factor of \boutT makes it difficult to use in some environments (with concealed/narrow outlets).

\begin{figure}[t]%
\centering
    \begin{subfigure}{0.43\columnwidth}
        \centering
        \includegraphics[width=\textwidth]{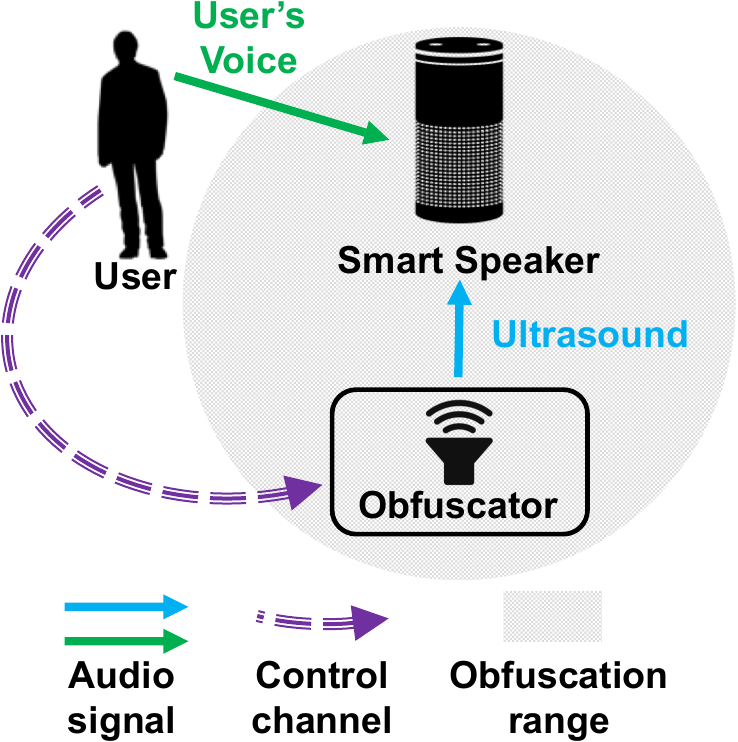}
        \caption{High-level overview}
        \label{fig:obf-overview}%
    \end{subfigure}
    \hfill
    \begin{subfigure}{0.5\columnwidth}
        \centering
        \includegraphics[width=\textwidth]{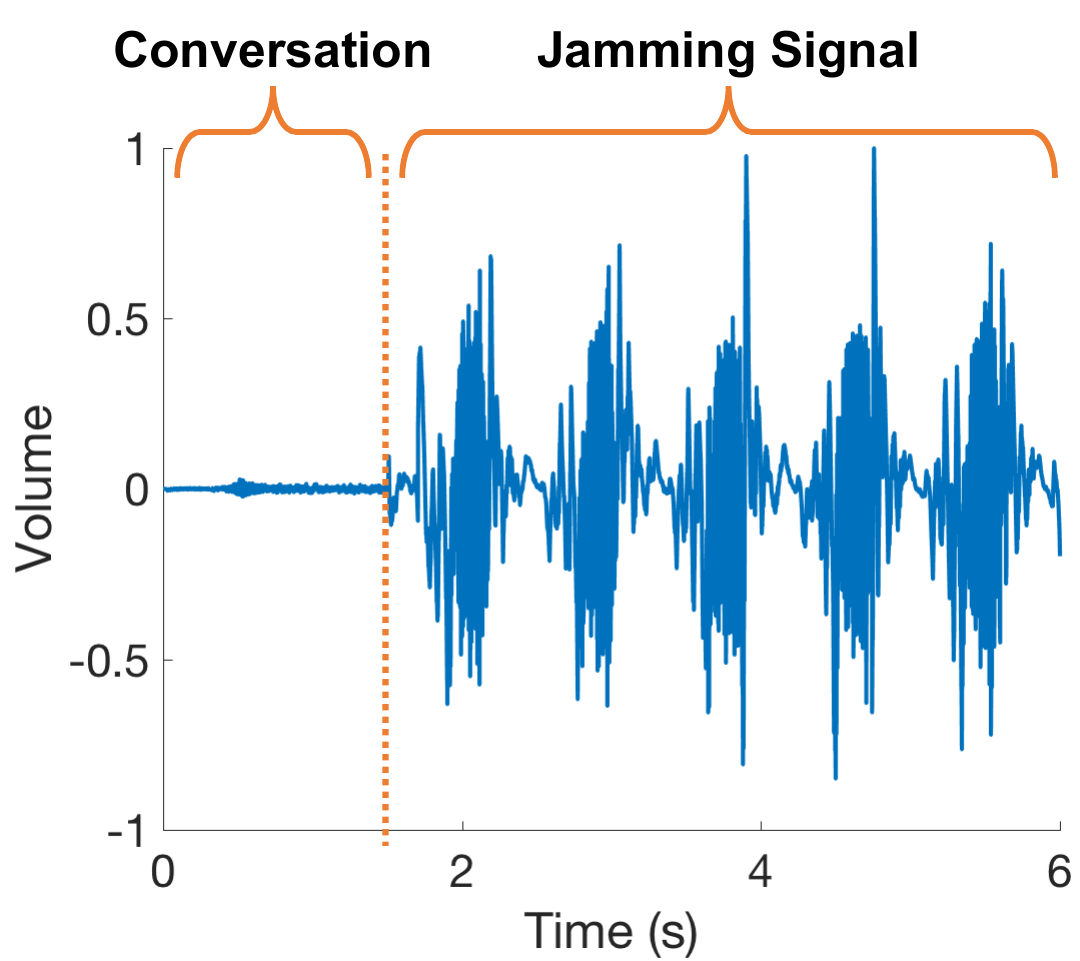}
        \caption{Microphone received signal}
        \label{fig:obf-signal}%
    \end{subfigure}
    \caption{The system design of the \obf probe.}
    \label{fig:obf-design}
\end{figure}

\vspace{1mm}
\noindent{\textbf{3. {\obfT}:}} This probe targets the microphone of the \va (Figure~\ref{fig:jam}). \obf generates {\em inaudible} ultrasound to deafen the microphone of the \va when the user needs privacy protection (Figure~\ref{fig:obf-overview}). Using a remote control, users are able to engage/disengage the probe without having to physically interact with it. When disengaging the jamming, the user can \textit{immediately} interact with the \va. Due to non-linearities in off-the-shelf microphones' power and diaphragm~\cite{roy2018inaudible,roy2018backdoor,zhang2017dolphinattack,gao2018traversing}, \obf creates high-power, human-inaudible noise at these microphones but does not affect its operation. Figure~\ref{fig:obf-signal} shows the captured signals from a commodity microphone before and after \obf is engaged. Before jamming is invoked, the microphone records a conversation, which is audible at playback. After engaging \obf, the ultrasound jamming signal is recorded at the microphone and completely overwhelms the conversation's signal. The circuitry of \obf includes a remote-controlled DC power supply, an ultrasound generator, and a horn speaker that emits the ultrasound signal. 

\textcolor{black}{The design of \obf utilizes a jamming signal with randomized tones at the ultrasound frequency range, which manifest as randomized tones at the audible range. Theoretically, a determined \va manufacturer can attempt to filter these tones at the expense of a degraded speech signal; such degradation might result in a deteriorated performance of wake word detection, which hinders the utility of the \va. Our experiments show that the jamming from \obf is effective at blocking the wake word detection.}

\subsubsection{Design Evolution}
\label{evolution}

We explored different design options for the prototype that houses the circuitry. A challenge in prototyping \obf was the footprint of the circuitry. Additionally, horn speakers are bulky, and reducing their size inhibits their efficacy. Our design process started with a search for a privacy metaphor, one that creates the perception of privacy control for the users. Our initial prototype was based on a ``cage'' metaphor. Here, the \obf probe is housed in a cage-like structure with a door, and the \va is placed within the cage. When the user closes the cage door, \obf generates the ultrasound obfuscation signal to prevent the \va from listening. The user has to manually open the door to disable obfuscation and communicate with the \va. Closing the door "locks'' the device in a cage, providing a user with a perception that the device is not active and their space is private. 

\begin{figure}[!b]%
\centering
    \begin{subfigure}{0.305\columnwidth}
        \centering
        \includegraphics[width=\textwidth]{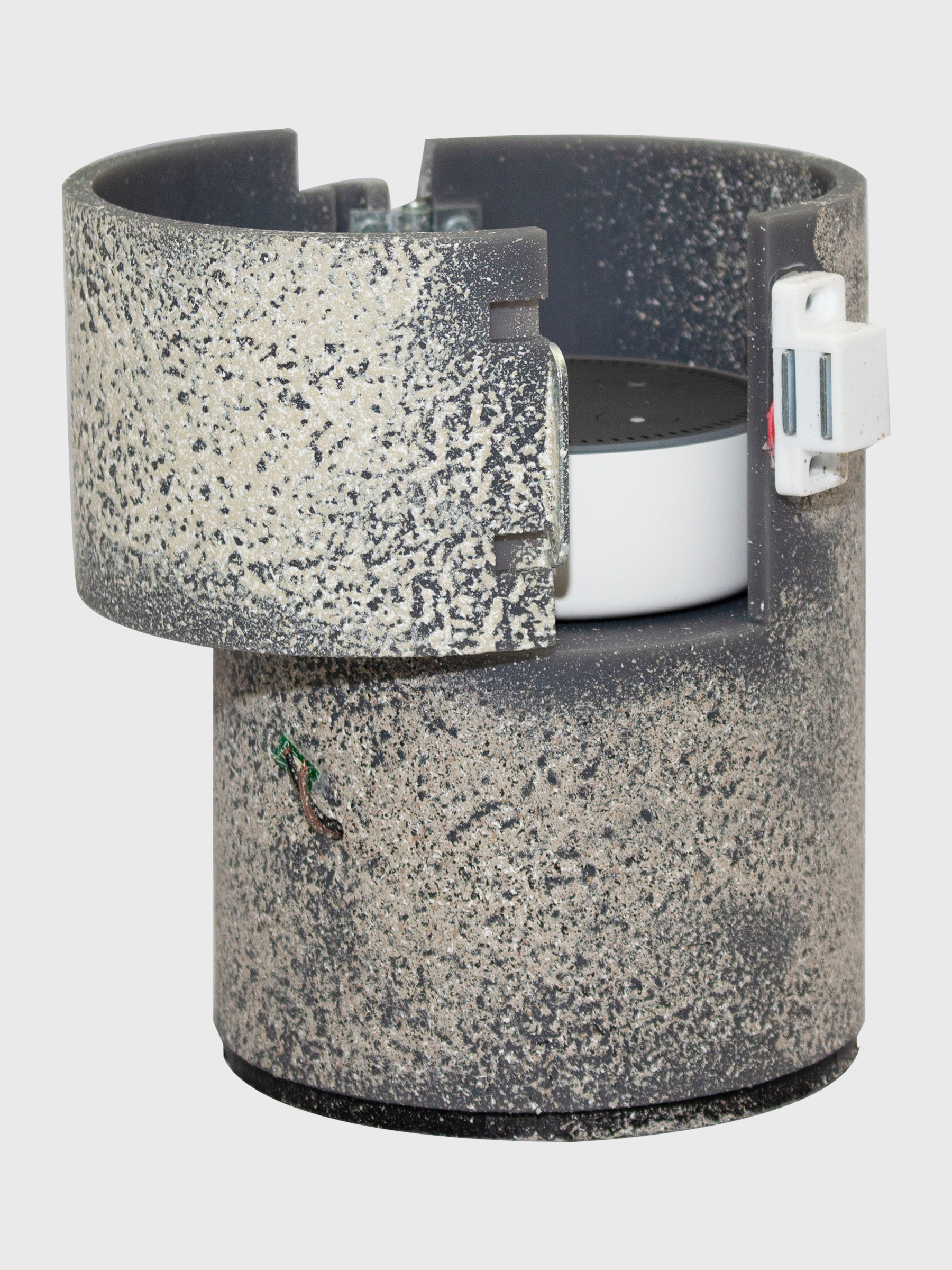}
        \caption{V.1}
        \label{fig:v1}%
    \end{subfigure}
    \begin{subfigure}{0.339\columnwidth}
        \centering
        \includegraphics[width=\textwidth]{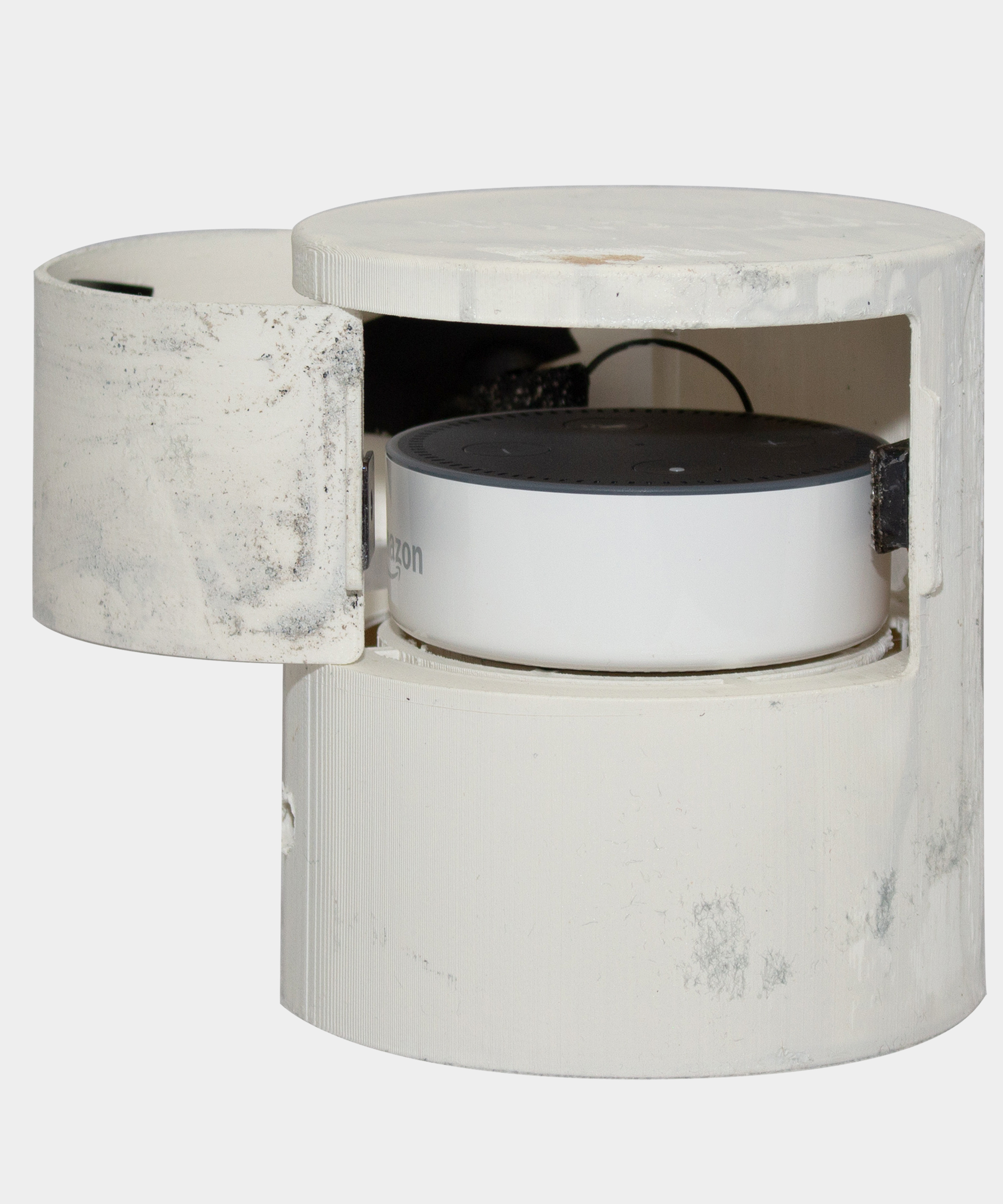}
        \caption{V.2}
        \label{fig:v2}%
    \end{subfigure}
    \begin{subfigure}{0.305\columnwidth}
        \centering
        \includegraphics[width=\textwidth]{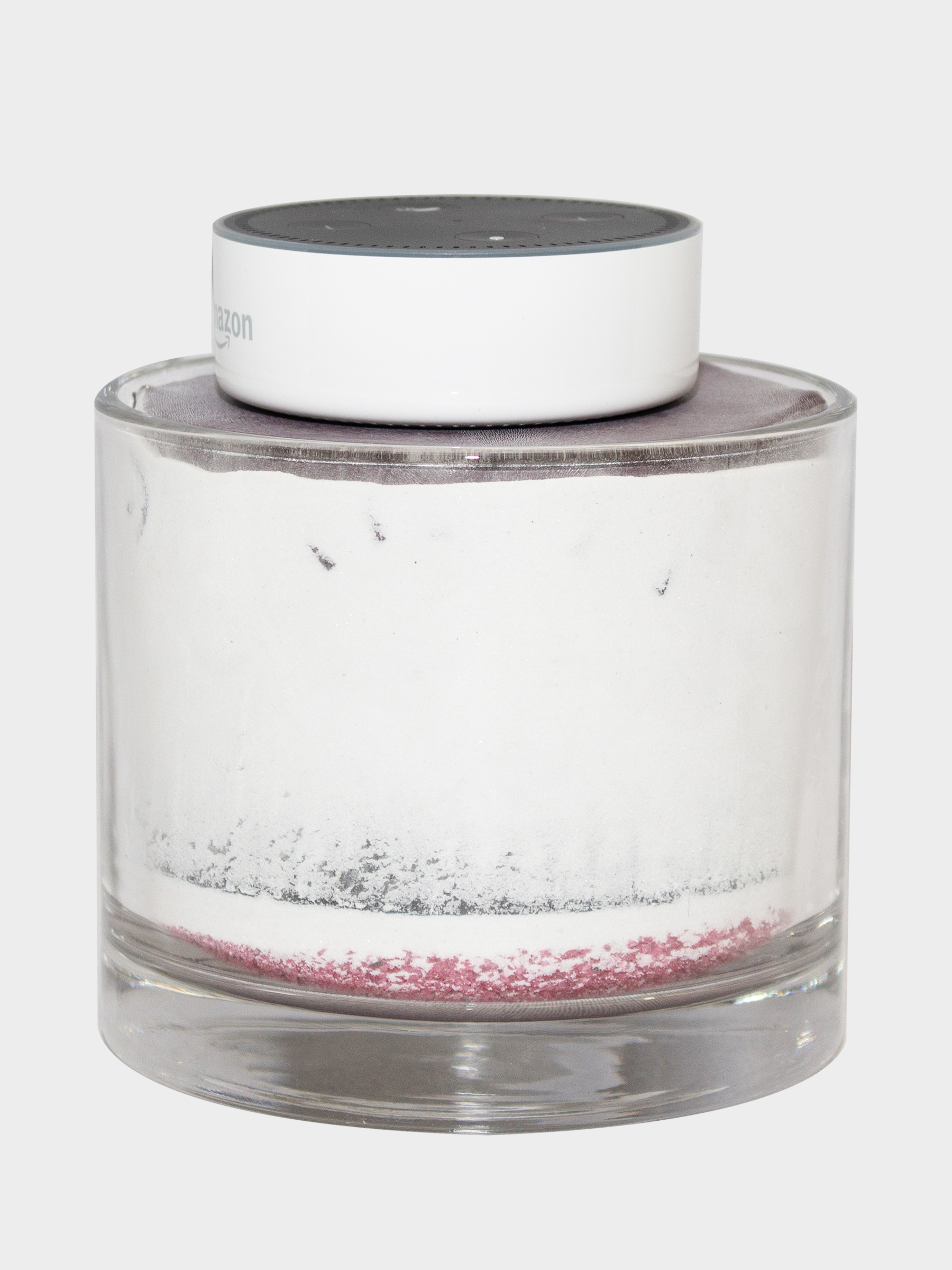}
        \caption{V.3}
        \label{fig:v3}%
    \end{subfigure}
    \caption{The design evolution of the \obf probe.}
    \label{fig:design_evolution}
\end{figure}

The first version of the \obf probe followed the cage metaphor as a 3D printed cylinder (Figure~\ref{fig:v1}). The cylinder has two compartments; the lower chamber containing the circuit and the ultrasound speaker. The upper chamber has space for the \va as well as the door. The first version has a height of 15.5 cm and a diameter of 12 cm. We refined this design into a lighter and less conspicuous 3D-printed cylinder (Figure~\ref{fig:v2}) with a height of 13 cm and a diameter of 11 cm. This was the second version.

Based on pilot studies with 2 participants, we found both versions to be neither user-friendly nor fitting with home decor. Participants explicitly indicated that this design was not something they would want in their homes. Further, we observed that individuals did not associate with the privacy metaphor. First, they did not favor the idea of physically interacting with the prototype as it takes away the convenience of using a hands-free device. Second, covering the \va inside the cylinder deprives the users of the ability to observe the visual cue (refer Table~\ref{table:features}). This is a shortcoming of placing the \va within the probe. Finally, they thought that the actions of opening and closing the prototype door were conspicuous and would rattle others in the vicinity.

In the third version, we considered three aspects that the users were not fond of: physical interactions with the door, covering the \va, and the aesthetics\footnote{Aesthetics are subjective, and determining a good aesthetic for even a prototype is a challenging problem.}. The third version of the prototype (Figure~\ref{fig:v3}) features a platform-like solution, which addresses those shortcomings. This version has a glass cylinder that houses the circuit and is covered by decorative sand; its height is 11 cm, and its diameter is 12.5 cm. The platform, where the \va sits, is encased with synthetic leather. The user can engage/disengage the jamming signal via remote control, obviating the need for physical interaction. This version of the \obf probe follows a different privacy metaphor: ``virtual veil.'' By engaging the jamming signal, \obf creates a virtual privacy dome around the \va, preventing it from listening to the conversations. Our subsequent discussions and reflections about this version revealed that the open nature of the prototype might not enforce the privacy metaphor; users are less likely to perceive privacy control over the \va. Additionally, this version remains co-located with the \va, increasing its form factor. This is not ideal when the \va is concealed. 

The design search process led to our final prototype of the \obf probe, as shown in Figure~\ref{fig:jam}. We substantially reduced the form factor of the final version. The new prototype houses the same circuitry in a glass candle holder. The glass is filled with decorative sands and sealed with burgundy burlap. The user only needs to place the prototype next to the \va. This prototype is built using commonly found household artifacts, enabling it to fit in with the existing decor. 
The final prototype (henceforth our probe) packages the core functionality of the privacy probe: a jamming device that enforces the privacy metaphor. We kept the prototype as simple and basic as possible to avoid making design decisions~\cite{buchenau2000experience} that influence our findings. In our study, we use the prototype to elicit participants' reflections about what design elements are missing and need to be introduced.

\subsection{In-home User Study}

We recruited 24 families (including single individuals) within a 15-mile radius of \varun{the UW-Madison campus}, utilizing the university mailing list, over two phases. Our first phase, in 2018, included 13 interviews, while the second phase (10 months later, in 2019) included 11 interviews. We use a 2-phased approach to obtain results from a wider variety of end-users; we wished to interview both unaware users (in phase 1) and those familiar with media reports of privacy violations induced by \va, at the time of the interviews (in phase 2). We chose to perform shorter and focused interviews as opposed to longer studies (such as diary studies); the tech probe approach allowed us to capture our many goals related to capturing baseline privacy perceptions, introduce the privacy priming, and gain reflection upon interacting with the interventions. 

The results reported in the paper are based on interviews with 30 participants ($P_1-P_{30}$) from these 24 interviews\footnote{Some households had more than one participant.}. Our data coding and analysis started immediately and took place simultaneously with data collection, enabling us to monitor the emergence of new codes and themes and determine saturation. We reached saturation by the 18$^{th}$ interview and collected data from 6 more households to assess how perceptions evolve with time. Our approach exhibits several limitations, the most important of which is the sampling of a relatively (a) {\em ethnically} homogeneous and (b) educated population; the reported results are less likely to generalize to another population of users. We sought to recruit participants with different backgrounds in age, education, and technological proficiency. Our participant pool comprised 15 males and 15 females. The youngest participant was 12 years old, while the oldest was 67 (with a mean age of 37.4 and a standard deviation of 13.9). The occupations of the participants ranged from students to faculty. The wide spectrum in age and profession enables us to gain feedback from a pool with varied technical knowledge and awareness and offers a breadth of experiences and backgrounds that are useful to analyze user interactions with the interventions. 

We conducted all interviews at the participants' homes at a time of their convenience. Each interview lasted 90 minutes on average, and the participants were compensated for their time (\$40 per study). The study protocol was approved by our Institutional Review Board. Each interview consisted of three stages, which we elaborate on below.

\vspace{1mm}
\noindent{\bf 1. Environment Exploration:} The interview began with the participants providing a brief tour of their home. Emphasis was placed on the rooms with \vas. Then, the interviewer and the participants convened in the room with the frequently used \va so as to simulate a common usage scenario. After obtaining informed consent, 
the interviewer \varun{first asked the participants to interact with their \va to ensure that it was operating as expected. This was followed by questioning participants about their knowledge/understanding of how \vas operate.} Then, the interviewer asked more detailed questions about the \va's role in the participant's life. The questions focused on frequency, duration, and the purpose of usage. Also, the questions covered the conversations and activities participants have around their \vas. Then, the interviewer inquired about the participant's degree of trust in these devices (in terms of the potential for their conversations to be recorded) and trust in their manufacturers and hypothetical third parties (with whom the recordings might be shared/leaked). The interviewer asked whether individuals have read the news or heard anecdotes about unexpected or undesirable behaviors by the \vas. These questions created the appropriate context to discuss privacy-preserving probes; while our follow-up questions are capable of biasing the participants, we believe that they are essential in creating the right environment to discuss the ambiguous space of privacy issues surrounding \vas.

\vspace{1mm}
\noindent{\bf 2. Interaction with Probes:} In a randomly generated order, the interviewer briefly introduced \varun{the probes and explained their capabilities to the participants}. The participants were given time to familiarize themselves with the probe and set it up (\ie reorganize their existing layout, if needed, to find a suitable location to utilize the probe). \varun{If this was not possible in the room where the interview was occurring, the interviewer and participants discussed why this was the case and moved to a more convenient location with a \va, should one exist.} By setting up the probe themselves, we expected the participants to gain greater familiarity with its operation and various other nuances (which we discuss later). The random ordering of probes across participants helped to reduce ordering effects. In settings with families, the interviewer asked different family members to interact with the probe individually (in the presence of other members). After setting up the probe, the interviewer asked the participant to issue voice commands after engaging/disengaging the probe. At each step, the interviewer probed the participant about their level of comfort with the probe and how it impacts the usability of their \va. The participants were encouraged to envision future use-cases for each probe and stress-test the probe's functionality. After the interaction with each probe, the interviewer inquired about the participant's level of trust in the probe. Based on the nature of the response, the interviewer asked several follow-up questions to determine reasons for high/low levels of trust. The interviewer proceeded to discuss perceived privacy control, trust level, convenience, and aesthetics of the probes. \varun{On average, users interacted with each probe for approximately 20 minutes\footnote{From our experience, the users were able to familiarize themselves with the mode of operation and installation of these probes in this timeframe.}.} These interview questions were designed to elicit critical reflections -- the primary aim of the tech probe study.

\vspace{1mm}
\noindent{\bf 3. Concluding Discussions:} The interviewer engaged the participants in an open-ended discussion about the probes and their impact on their privacy. \varun{The interviewer allowed the participants to hold and observe our probes before answering any other questions related to the study. Finally, the interviewer compensated the participants for their time and effort.}

We recorded the interviews, resulting in over 30 hours of recordings, and took photographs of (a) the probes in action and (b) areas where the \va is typically used. We then transcribed, coded, and analyzed the interviews using a Grounded Theory approach~\cite{charmaz2012qualitative, glaser2017discovery}. The coding was performed with two coders working independently. Our coders were in moderate agreement, with a Cohen's Kappa ($\kappa$) of 0.57~\cite{viera2005understanding}. We started the analysis with an open-coding stage to identify more than 200 informal codes that define critical phenomena in the interview transcripts. Using these informal codes, we extracted recurrent themes within the transcripts and converged on a set of 88 formal codes. We further refined the formal codes into 15 axial codes. We organized the codes into three major themes as summarized in Table~\ref{themes}. \varun{We believe that the value of the agreement is acceptable for our study, based on previous research~\cite{mchugh2012interrater}. Following common practices in qualitative coding~\cite{belur2021interrater}, disagreements were discussed by the coders, followed by code reconciliation, resulting in an updated codebook.}



\section{Observations}
\label{sec:findings}

\begin{table}[t]
\small
\label{table:themes}
\begin{center}
\begin{tabular}{ p{8cm}}
\toprule
{\bf Attitudes towards Smart Speaker} \\ 
\midrule
Characterizes the user's (a) nature and awareness, (b) technological know-how, and (c) trust in \va manufacturer. \\ 
\toprule
{\bf Attitudes towards Probe}\\ 
\midrule
Characterizes the user's (a) interaction preference, (b) comfort-levels with regards to usage, (c) long-term technological preferences, (d) trust attitude towards probe, and (e) aesthetics and physical footprint preference.\\ 
\toprule
{\bf Utility of Probes}\\ 
\midrule
Characterizes the user's preference with respect to probe's (a) multi-functionality, (b) cost, (c) ability to provide fine-grained control, and (d) mode of operation \ie proactive vs. reactive,  \\ 
\bottomrule
\end{tabular}
\end{center}
\caption{Summary of the extracted themes.}
\label{themes}
\end{table}

In this section, we discuss the central themes that emerged from our analysis. In summary, we found that: (1) participants were reluctant in sacrificing the convenience associated with \vas; hands-free interaction was most preferred, and physical interaction was seen as being not ideal; (2) participants expected bolt-on interventions with existing household decor and to offer cues informing them of the state of both the probe and the \va; and (3) participants had a preference for multi-functionality and fine-grained control (per-user and per-device). \varun{Several of the observations we make have been reported earlier~\cite{abdi2019more,laualexa,huang2020amazon,malkin2019privacy}. Our work re-affirms them and shows that the sample used for the rest of the analysis reveals consistent perceptions as previous work\footnote{These findings resulted from our observations in 2018, predating many of the works cited here.}.}

\subsection{User Attitudes regarding Smart Speakers}

\vspace{0.5mm}\noindent{\bf 1. Types of Users:} Through our study, we identified two types of users: (a) {\em casual users} who utilize their \vas for setting alarms, asking questions, etc., and (b) {\em power users} who have integrated the \vas with other devices in their homes (such as smart lights, house monitoring systems, etc.). We also observed that most participants in our first interview phase were casual users, and a majority of those in the second phase were power users. This phenomenon could be based on the pervasive availability of various smart home devices. We observed that power users (and those in the second phase) were also more familiar with passive privacy violations and with the potential for active violations. We observed that power users were more willing to adapt privacy-preserving interventions as their households were more tightly integrated with the \va. We also observed that a majority of the participants did not change their conversations around the \vas, but a small minority reported feeling conscious of having discussions around them. \varun{Similar observations were made in recent works studying the privacy perceptions/attitudes of smart speaker users~\cite{abdi2019more,laualexa,huang2020amazon,malkin2019privacy}.}

\vspace{0.5mm}\noindent{\bf 2. Understanding of Smart Speaker Operation:} A minority of the participants was unaware of how \va's operate, \ie they were unaware that their voice commands were processed off-site. Participant $P_5$, for example, believed that the \vas did \textit{``some local learning but also some more... I think at some point people were involved in [the processing]... I think there's an automated learning that occurs to adjust itself to the household, right?''} \varun{Abdi \etal reported similar observations about users having incomplete mental models of the smart home personal assistants~\cite{abdi2019more}.}

\vspace{0.5mm}\noindent{\bf 3. Trust in Device Manufacturers:} Our participant pool includes fractions (a) that believed that these organizations could be trusted, (b) that believed that some manufacturers were not in the business of collecting personal information and can be trusted, (c) that trusted the manufacturers, but believed that any information collected could be leaked, and (d) that trusted the manufacturers as long as there is personal utility gained from disclosing said information. A recurrent theme was participants' comfort in being recorded because they believed they were part of a large pool of \va users. Participant $P_{10}$ explains, \textit{``I mean we're not planning any nefarious capers... like we're very boring people and therefore nothing that we're talking about would be of interest to anyone on the other end of [the \va].''} \varun{Other studies have also studied user's trust in the device manufacturers and have reached similar conclusions~\cite{laualexa,huang2020amazon}.}

\subsection{User Attitudes Regardingt Probes}

\vspace{0.5mm}\noindent{\bf 1. State of Operation:} Participants believed that the current designs of the probes make it too inconvenient to use the \va. They state that using them makes the interaction with a \va a two-phase procedure: first, check the state of the probe (engaged vs. not) and disengage if necessary, and then interact with the \va. Some participants stated that the probes added a {\em mental burden} in terms of remembering its state. Participant $P_9$ said: {\em ``when you were to power it off say how do you distinguish that state [when it has no power when using \boutT] from a wake word doing nothing, like I don't know I unmute this right now ... it looks the same.''}

\vspace{0.5mm}\noindent{\bf 2. Ergonomics:} Participants were comfortable with the {\em usability} of the probes. They were easy to set up and use, and the time taken for the probes to activate is acceptable (almost instantaneous in all cases). However, participants expressed dissatisfaction at the longer boot-up times induced by \boutT. For example, $P_8$ stated,\textit{``I would find it especially irritating.''} Participants suggested that technologies such as \obfT that, when disengaged, make the \va {\em immediately} available were {\em ideal}. Some participants were concerned about the generalizability of \obfT. They believed that the technology is specific to their \vas, and would not extend to future \vas or \vas made by other vendors. Participants were comfortable using a remote control but felt that their homes have many remotes that could be easily misplaced. When proposing the addition of another remote, $P_5$ exclaimed, \textit{``they're all over the place, so many remotes! We can't have another remote.''} Some participants suggested moving intervention control to a mobile phone app. 

\vspace{0.5mm}\noindent{\bf 3. Trust in Bolt-On Interventions:} Finally, participants trust our bolt-on probes more than the built-in mute button. However, participants suggested that trust in a bolt-on intervention would be low if it came from the device manufacturer or any organization that had a similar business model. Participant $P_{13}$ recommended \textit{``a competing company or just a general company that seems like they're like honest''} could develop the interventions. Participants suggested that bolt-on interventions were easier to debug and were easier to understand. However, participants feel that purchasing one bolt-on intervention for every \va would be expensive.

\begin{figure}
 \centering
  \includegraphics[width=\columnwidth]{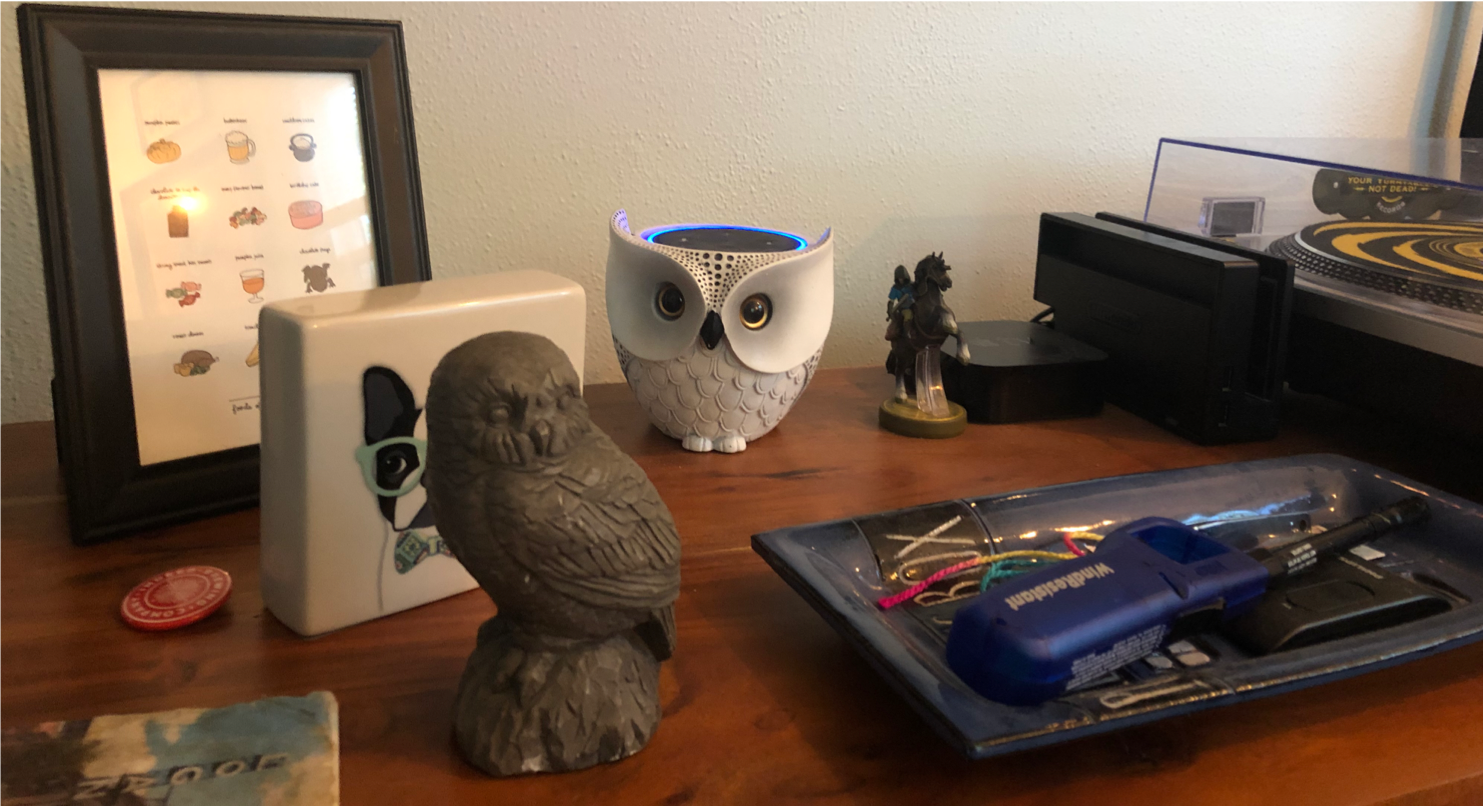}
  \caption{The placement of an Amazon Echo Dot inside an owl-shaped holder in one of the households.\vspace*{-6mm}}
  \label{fig:placement}
\end{figure} 

\vspace{0.5mm}\noindent{\bf 4. Physical Footprint:} Participants were concerned with the physical footprint of our probes. While \vas were electronic devices, participants often associate them with decorative items (Figure~\ref{fig:placement}) and invest effort in determining where these devices should be placed. A common example of a description about the \obfT solution we received was $P_2$'s description: \textit{``a piling on of devices.''}  Some participants found it difficult to reorganize other items around the \va to facilitate the probe. Additionally, some participants prefer to conceal their outlets, and \boutT-like interventions were inconvenient in such scenarios. Participants were uncomfortable with interventions that involve additional wires (as in the case of \obfT). \varun{Similar observations were made by Pateman \etal~\cite{pateman2018role} in the context of the adoption of wearable devices.}

\subsection{Utility of the Probes}

\vspace{0.5mm}\noindent{\bf 1. Damage to the Environment:} Participants were concerned that \obfT would cause harm to nearby animals; questions we received upon presenting the \obfT were often like $P_2$'s, \textit{``is [this] going to ... make my dog crazy?''} While we did not observe any agitation/discomfort, the participants suggested that their pets could perceive the ultrasound signals and were not bothered. Additionally, participants were concerned about exposing their \vas to ultrasound for a prolonged period of time\footnote{A detailed study is needed to understand the impact of ultrasound on electronic devices.}. 

\vspace{0.5mm}\noindent{\bf 2. Cost and Multi-Functionality:} Cost was repeatedly discussed; participants suggested that the cost of the interventions should not exceed the cost of the \va. Some participants received their \vas as gifts. Consequently, they were unable to establish a value for an intervention; $P_6$ states, \textit{``that's a really interesting question in the sense that I didn't pay for this in the first place. Maybe that's also another reason that I don't have much investment in using this in general.''} On the other end of the investment spectrum, we observed that participants who owned multiple smart devices were invested in safeguarding their privacy and were willing to adopt interventions independent of the cost. Participant $P_6$, who had previously stopped using their \vas due to privacy concerns, even stated that they would consider using their device once more given that the interventions were \textit{``cheap... I think would have to rival that remote plug-in cost right because ... it has to be like a cheap utility ... or a cheap accessory like that.''} \obfT could be used in a proactive way \ie always-on, or in a reactive way \ie use when needed. Participants felt that a reactive approach, though tedious, would be easier to understand. Participants also believed that cost could be justified if the intervention provided multiple features. This could be achieved by integrating the design of \obfT with other home decors, such as lamps, lights, clocks, radios. 

\varun{\vspace{0.5mm}{\bf 3. Multi-user and Multi-device Environments:} The final observation we make is an extension to multi-user and multi-device environments; we observed that in some households, some participants preferred to utilize the intervention more than others. Also, different types of users might exhibit different privacy requirements when interacting with \vas. In such scenarios, they desired customized usage profiles based on their requirements \ie \xspace {\em access control per-user}. Recent research has also indicated the need for access control flexibility in multi-user smart homes~\cite{zeng2019understanding}.} Another observation we make is that some participants preferred to have one intervention (like \obfT) being used to preserve privacy against a wide range of \va-like devices. In such scenarios, {\em access control per-device} was desired. Based on the current design of the \obfT prototype, meeting both these requirements is challenging and requires further research. \varun{One research direction to make access control per-user more feasible is establishing default privacy options depending on the expected user privacy profiles~\cite{abdi2021privacy}, such as owner vs. visitor.}

\section{Design Implications}
\label{sec:design}

\begin{table}[!ht]
\small
\begin{center}
\begin{tabular}{ p{8.1cm}}
\toprule
{\bf Concrete Recommendations}\\
\toprule
{\bf 1. Aesthetics:} The interventions should be offered in different forms, shapes, and colors to fit within people's decors and furniture.
 \\ 
\midrule
{\bf 2. Physical Footprint:} The footprint of the intervention should be small enough to not force a reorganization of the layout of the owner's house. \\ 
\midrule
{\bf 3. Multi-Functionality:} The intervention is better when providing additional functionality (such as a clock) to reduce its footprint and integrate better with home decor. \\ 
\midrule
{\bf 4. Ease of Deployment and Understanding:} Battery-powered interventions are easier to deploy. \\ 
\midrule
{\bf 5. Ease of Understanding:} A proper understanding of the privacy metaphor improves the adoption of interventions. \\ 
\midrule
{\bf 6. Trust in Technology:} Trustworthy interventions are bolt-on, not network connected, designed by a different trustworthy organization, and pose no additional risk. \\ 
\midrule
{\bf 7. Mode of Interaction:} Using the intervention should not change the interaction with the \va. Hands-free interaction is preferable. 
\\ 
\midrule
{\bf 8. Informative Cues:} Interventions should offer cues that communicate their state. Visual, auditory, or text cues might be applicable depending on the deployment. \\ 
\toprule
{\bf 9. Cost:} The intervention should cost less than the \va. \\ 
\midrule
{\bf 10. Fine-grained Privacy Control:} The intervention can offer per-user and per-\va privacy controls.  \\ 
\midrule
{\bf 11. Awareness:} Awareness of privacy violations increases trust in intervention designers.  \\ 
\bottomrule
\end{tabular}
\end{center}
\caption{Summary of the identified design guidelines.}
\label{tab:findings}
\end{table}

Based on the findings from \S~\ref{sec:findings}, we make concrete recommendations (based on our findings) on how to design privacy-preserving interventions. The design recommendations are along axes specified in Table~\ref{tab:findings}.

\noindent{\bf 1. Aesthetics:} We observed the aesthetics of the privacy interventions to be an important issue for our participants. Participants preferred the interventions to match their individual decorating styles (one example is shown in Figure~\ref{fig:placement}). Many participants suggested that the intervention should come in different forms, shapes, and colors, enabling easier integration within their home decor. As individual tastes vary widely, devising a one-fits-all design is challenging. {\em One possible approach is to explore different design options for different types of users, including shapes, forms, colors, and material}. This approach has been successful with \vas, where participants feel comfortable with the aesthetic of the \va. For example, Amazon has four variants of their Echo featuring combinations of forms and fabric colors. 

\noindent{\bf 2. Physical Footprint:} Since the \vas we considered were small and compact, participants preferred a similar physical footprint for the interventions. Participants expressed concerns regarding the size of both \boutT and \obfT, enquiring if a similar functionality could be achieved with a smaller probe. They believed that using \obfT (which needs to be proximate to the \va) requires them to significantly reorganize their existing home decor layout. While the form factor of \boutT can be reduced trivially, doing so for \obfT is challenging; the size and shape of the horn speaker in our current probe were chosen to ensure maximum ultrasound distribution and coverage. Extending such a design to (newer) \vas that are larger, or have a different orientation for the microphone inputs, will require rethinking the design and form factor. In summary, interventions that require proximity to the \va need to be designed such that their form factor is comparable to the \va. To achieve such a design, {\em one recommendation is to design the \obfT-style intervention as a stand (upon which the \va can be placed), or as an artifact that can be placed above the \va.} In both designs, the intervention will generate a veil of ultrasound around the entire \va (similar to the horn speaker case that we had designed and evaluated). 

\noindent{\bf 3. Multi-Functionality:} Closely tied to the aesthetics, participants indicated preference toward an intervention (specifically \obfT) that offered features beyond privacy-preservation. They suggested that the \obfT intervention could be combined with other household artifacts, such as a lamp, radio, clock, which would further improve adoption. Additionally, {\em multi-functionality provides an alternative avenue for customizing the probe, making it easier to integrate with existing household decoration}. Such products alleviate the social stigma of being labeled as overly privacy-conscious; such stigma is another reason why the adoption of privacy-preserving interventions is currently low. 

\noindent{\bf 4. Ease of Deployment:} Participants state that they prefer having a solution that is easy to deploy in their homes; the biggest impediment to any intervention similar to \boutT is its requirement for an outlet. Many participants preferred to conceal the interventions' wiring, and the nearby outlets can be hard to reach. Attaching \boutT to wall outlets, even once, requires considerable re-positioning of other devices and their wires. Attaching \obfT would require an additional outlet, which is not always readily available. One na\"ive solution would be to split the outlet among multiple devices. Participants suggested that an \obfT design capable of operating on batteries would be more preferred, even if this required periodic replacement. 

\begin{tcolorbox}
\noindent{\bf Recommendation:} Combining the above four observations, we recommend designing interventions in one of two forms: (a) a stand to hold the \va, or (b) a sleeve for the \va (refer Figures~\ref{fig:amazon1} and~\ref{fig:amazon2}). Based on some preliminary analysis, we observe that there is a demand for such artifacts based on our analysis of reviews for such products, and we believe such designs would promote adoption. Since the intervention is not operational in an {\em always-on} mode, it may be battery-powered — doing away with the requirement for an outlet.
\end{tcolorbox}



\noindent{\bf 5. Ease of Understanding:} All participants were able to easily grasp the metaphor associated with \boutT, but the technology behind \obfT proved complicated for some; some users were unfamiliar with how ultrasound induces a deafening effect. Thus, interventions whose operation is easy to explain may be preferred. This is particularly the case because, while \obfT is easy to use once deployed, debugging it may pose problems for users who lack a proper understanding of its operation. We also believe that understanding the detriments (if any) of ultrasound towards humans, animals, and other electronics may put users at ease.

\noindent{\bf 6. Trust in the Technology:} Participants were more comfortable with technologies that they believe will survive the ``test of time,'' \ie be useful for \va models in the future. As discussed earlier, trust also stems from knowing that the interventions do not pose any additional risk. Specifically, we observed that (a) participants wanted to know about any detriments introduced by the interventions, such as potential damage to the \va by frequently disconnecting it from its power source or subjecting it to ultrasound; and (b) our current interventions are not network connected and do not present the same risks as the \vas. Finally, participants preferred our bolt-on interventions as opposed to the built-in interventions as they were designed by an organization they trusted (more than the \va manufacturers). 

\begin{tcolorbox}
\noindent{\bf Recommendation:} Combining the two points stated above, a concrete design recommendation is to communicate the science behind the operation of the \boutT-style intervention with a more relatable metaphor or through an interactive demonstration of the intervention's operation. By doing so, we are able to provide more intuition on failure scenarios, which can enable more efficient debugging. This process also assuages any fears related to \va damage or possible harm to nearby entities (such as pets). 
\end{tcolorbox}

\noindent{\bf 7. Mode of Interaction:} We observed that participants placed a high value on the {\em convenience} of using \vas, which they are not willing to compromise. Thus, interventions that, when engaged, delay the \va operation (as in the case of \boutT) are not preferred (even though \boutT provably preserves privacy, and its mode of operation is very easy to understand). Additionally, any form of physical interaction, be it using remotes or buttons, is far from ideal; some participants expressed preference toward using an app on their smartphones. 

\begin{tcolorbox}
\noindent{\bf Recommendation:} We believe that future interventions must be designed so as to have minimal disruption to the convenience of the use of these systems. An ideal design would have a voice interface that allows the user to control it as they control their \vas. However, such an always-on and listening privacy-preserving solution can have the same pitfalls as \vas, and they must be designed in a manner that does not erode user trust; the mechanism to provide privacy (via a voice-interface) must not become a mechanism for exfiltrating sensitive user conversation (as such a mechanism may require to be network connected). For example, they can lack a network interface to provide the users with hard privacy assurances. Another issue that may arise with voice-activated interventions is erroneous activations; understanding how this can be minimized requires additional research. 
\end{tcolorbox}
 
 
\noindent{\bf 8. Informative Cues:} As stated earlier, some participants concealed their \vas and would prefer concealing their interventions as well. Some participants take this notion to the extreme; they believe that any electronic device that does not provide extensive visual information should be concealed. Thus, visual cues are not ideal in all situations. Additionally, participants suggested that the {\em red} light on the \boutT intervention suggested that the intervention was broken, as opposed to indicating the state of the intervention. Interacting with the \va when the intervention is enabled helps users determine the state of the \va (operational vs. not), but such an approach is reactive. Participants indicated a preference for a {\em proactive} approach. 

\begin{tcolorbox}
\noindent{\bf Recommendation:} We propose two recommendations for such settings: (a) the state of the intervention (i.e., engaged vs. disengaged) by communicating to a device that is more optimally placed for being viewed (such as a TV) — this can be done using some form of a closed network connection between the TV and the device via Bluetooth, or (b) the intervention provide auditory cues, where the \obfT-style intervention can announce using speech or text that the \va is inactive when users try to activate it.
\end{tcolorbox} 


\noindent{\bf 9. Cost:} Another factor that impacts adoption is the cost of the \va. A large fraction of our participants owns smart plugs similar to \boutT, leading us to believe that such an intervention is affordable. However, the cost of prototyping \obfT was \$70, exceeding the cost of \vas (priced at approx. \$30). This cost includes the price of the commodity parts needed to construct the probe. Participants believe that the cost of the intervention should not exceed the cost of the \va; this is especially true if the intervention can provide privacy protection against a single \va. We believe that if such an intervention would be adopted widely, the production costs could be amortized (and thus have no concrete recommendation to make with regards to minimizing cost). Additionally, understanding the engineering requirements to design an \obfT-like intervention that provides privacy against various \vas located at different parts of a home requires independent research. 


\noindent{\bf 10. Fine-grained Privacy Control:} Several households owned more than a single \va, and they had members with different (and potentially conflicting) privacy requirements. Thus, we believe that there is a requirement for (a) fine-grained control \textit{per user}, and (b) fine-grained control \textit{per \va}. For the latter, a na\"ive solution would be to deploy one intervention per \va, but depending on the cost per intervention, such a solution may not scale. Providing per-user control is a more challenging problem; it requires understanding how disparate the privacy requirements are, how frequently users are utilizing a \va together, and how to mitigate conflicts should they arise.

\begin{tcolorbox}
\noindent{\bf Recommendation:} An ideal design would provide privacy protection for more than one \va. This design could be conceptualized as smaller interventions co-located with the \vas but controlled centrally (through some form of closed network).
\end{tcolorbox}

\noindent{\bf 11. Effect of Awareness:} Based on our interview questions, we observed the following trend amidst the participants of our interview phases. The participants of our second phase are more concerned about the potential privacy threats from the \vas (in comparison to the participants of the first phase, who are also concerned). This concern stems from increased awareness, recent \va mishaps, erroneous code used in them, and immoral practices by device manufacturers. Based on our discussion, we observed that participants believe that these issues are not being seriously audited by the device manufacturers. Discussing various loopholes that can be implemented in the built-in interventions in the status quo (\ie local wake word processing and the mute button) also increased participants' awareness. 


\subsection{Consolidated Recommendation}

We consolidate the design recommendations based on the aforementioned discussion and provide concrete design guidelines.

\varun{\noindent{\bf Aesthetics, Utility, and Accessibility:} \obfT-like interventions should be incorporated in accessories that users are already adopting, such as stands and holders (the ``owl'' shown in Fig~\ref{fig:placement}). Since the completion of our work, we have seen an emergence of a market for such accessories. Additionally, the jamming device should be hidden within a device that is multi-functional, privacy being the secondary functionality. Further, the jamming device should be always-on; the user can access the \va through hands-free interactions, such as gesture-based interaction through wireless sensing. A chime can be played to indicate that the \va is currently active.} 

\varun{\noindent{\bf Cost \& Centralized Control:} \obfT-like jamming systems rely on directionality to enable their functionality. Thus, it is unclear if there can be {\em one} of such solutions for {\em multiple} \vas in a home environment. However, many such interventions can be controlled through a centralized interface, such as a single remote control or mobile phone app. Future research is required to better understand the requirements of such a control interface and design it.} 

\varun{\noindent{\bf Building User Trust:} To enhance trust in such interventions, video (or other forms of) presentations/materials can be made to indicate that current \vas are purported to exfiltrate home conversation through the use of the public internet. Once this is established, we can educate users of the fact that \obfT-like interventions are not connected to any network and consequently can not share sensitive (or any other) information. To further strengthen user belief in bolt-on solutions, end-users can be educated about issues with built-in solutions. Notably, make changes to any built-in solution after deployment requires device manufacturers to regularly and reliably share software updates. However, installing such updates is a challenging proposition to even tech-savvy users~\cite{sans}. Additionally, as the ecosystem of such \va devices is fast evolving, manufacturers will often not provide support to (a large volume of) \vas that were deployed in the past~\cite{iotinsecure}. Additionally, information about software updates (needed for built-in solutions) is not easily accessible, resulting in periods of privacy loss~\cite{foscam}.} 

\varun{\noindent{\bf Accessibility in a Multi-User Environment:} Since different users may have different privacy requirements, interventions may be designed to operate with different profiles, such as always-on versus selectively turned on. However, choosing the profile may require (a) explicit user interaction with the intervention, which may be inconvenient, or (b) using auxiliary hardware to identify the users~\cite{feng2017continuous}.}

\varun{\noindent{\bf Enhancing Awareness:} Finally, we recommend an on-boarding process that educates the users about the potential privacy threats from accidental/malicious activations, the technology underlying the operation of the intervention, and how to utilize the privacy controls. Such an on-boarding process can take place through voice prompts or an external app; it will increase the user's awareness of the privacy issues as well as improve the user's trust in the probe.}

\section{Related Work}

\noindent{\bf Privacy Perceptions:} The methodology of our study is most similar to Zheng \etal~\cite{zheng2018user}, and Kaaz \etal~\cite{kaaz2017understanding}. They attempt to understand the privacy perceptions of users living in homes with various IoT devices. Similar to our work, surveys are carried out in~\cite{choe2011living, mccreary2016contextual, brush2011home}, where the authors try to identify the various challenges associated with setting up and using these devices. Zeng \etal~\cite{zeng2017end} and Lau \etal~\cite{laualexa} study \va users' reasons for adoption through a combination of a detailed diary study and in-home interviews. Similar to this work, we observe that \va users are not privacy-conscious because of the lack of value they associate with their conversational data. Along a similar vein, Abdi \etal~\cite{abdi2019more} find that users have incomplete mental models of \vas. Similar to our work, they use this understanding to present design recommendations. Some of our findings are coherent with those of Malkin \etal~\cite{malkin2019privacy}, \eg participants are unaware that their conversations are being recorded and stored. 

We stress that the primary contribution of our work is {\em not} in ascertaining the privacy perceptions people have about \vas (as done in earlier studies). We wish to understand users' perceptions towards privacy-enhancing technologies and to use this insight to guide the design of both \vas and such technologies. 

\vspace{1mm}
\noindent{\bf Probe Design:} Prior research has investigated system-level solutions to these privacy threats. Feng \etal~\cite{feng2017continuous} propose continuous authentication as a mechanism to thwart privacy issues related to \vas. In our previous work, we propose using ultrasound jamming to address stealthy recording \etal~\cite{gao2018traversing}. In this work, we wish to validate the usability claims made by the above; consequently, we base our intervention design on the above proposals. 
\varun{The works of McMillan \etal~\cite{mcmillan2019designing} and Mhaidli \etal~\cite{mhaidli2020listen} provide hands-free alternatives. However, the introduction of a camera to measure gaze introduces privacy concerns. This also requires the user to be in the line of sight of the \va, which reduces its usability. Solutions based on pitch and volume~\cite{mhaidli2020listen} also suffer from similar proximity issues and fail to eliminate privacy violations due to accidental activations.}

The Alias project~\cite{alias} is designed to achieve similar goals to ours. This solution constantly plays noise through a small speaker placed atop the \va and stops the noise upon hearing a custom wake word. Their solution differs from \obf in two ways. First, the Alias intervention does not use ultrasound; the reduced form factor is achieved by not using horn speakers, which are crucial for transmitting ultrasound. 
Second, the Alias intervention obscures the visual cue provided by the \va; such a design is not preferred. Similarly, work by Chen \etal~\cite{chenwearable} designs a wearable intervention. \varun{Wearable solutions offer support in some scenarios, \eg mobile situations. However, they offer poor support for \va due to lack of proximity to the device. Our experiments with ultrasound-based jamming revealed that the direction of the jamming device and the distance to the \va impact its performance. Additionally, Chen \etal do not evaluate the user-related aspects of the intervention, such as user acceptance, aesthetics, and trust.}



\vspace{1mm}
\noindent{\bf Design Studies:} To safeguard privacy and security in the smart home, Zeng \etal~\cite{zeng2019understanding} prototyped a smart home app and evaluated its effectiveness through a month-long in-home user study with seven households; the users are assumed to be non-adversarial and cooperative. They used their findings to guide future designs for smart home applications. To achieve similar goals as ours, but for smart homes (as opposed to \vas), Yao \etal~\cite{yao2019defending} adopted a co-design approach and designed solutions with non-expert users. We borrow our study methodology from the work of Odom \etal~\cite{odom:2012}; technology probe studies serve multiple purposes related to designing, prototyping, and field testing the interventions.



\section{Conclusions}


We presented the design and prototyping of two privacy-preserving interventions: `\obfT' targeted at disabling recording at the microphones, and `\boutT' targeted at disabling power to the \va. We presented our findings from a technology probe study involving 24 households that interacted with our prototypes, aimed to gain a better understanding of this design space. Our study revealed several design dimensions for the design of privacy interventions for \vas, including multi-functionality, trustworthiness, cues, interaction mode, and ease of deployment.

\section*{Acknowledgments}

We would like to thank Christopher Little and Thomas Linden who helped with the interviews. We would also like to thank Mariam Fawaz who assisted with the photographs of the interventions, and Yilong Li who assisted with the fabrication of \obfT. Finally, we would like to thank the anonymous reviewers and our shepherd for their constructive feedback. Varun, Suman, and Kassem were supported in part through the following US NSF grants: CNS-1838733, CNS-1719336, CNS-1647152, CNS-1629833, CNS-1942014, and CNS-2003129 and an award from the US Department of Commerce with award number 70NANB21H043.

\newpage
\bibliographystyle{plain}
\bibliography{references}

\newpage
\appendix
\section*{Appendix}

\section{Formal Codes}
\label{appendix}

\begin{enumerate}
\item privacy awareness vs education = awareness function of education
\item privacy awareness vs education = not equivalent
\item user technical knowledge = high
\item user technical knowledge = medium
\item user technical knowledge = low
\item user technical knowledge = varies in home
\item user education level = high
\item user education level = medium
\item user education level = low
\item user has concern = yes listening
\item user has concern = yes recording
\item user has concern = yes other
\item user has concern = no
\item user trust large orgs = yes
\item user trust large orgs = case by case
\item user trust large orgs = no
\item user trust third party = yes
\item user trust third party = no
\item user type = power user
\item user type = simple user
\item user accepts listening if = choose over recording
\item user accepts listening if = machine only
\item user accepts recording if = utility
\item user solution choice = discard device
\item user solution choice = unplug device
\item user solution choice = mute
\item user solution choice = remote plug
\item user solution choice = obfuscator
\item user believes in intervention = maybe
\item user believes in intervention = no
\item va state listening = wake word only
\item va state listening = yes
\item va state recording = yes
\item va state recording = non human
\item va state issue attribution = bugs
\item va state issue attribution = unaware
\item va state data use = mundane
\item va state data use = nefarious
\item ecosystem factor = space for solution
\item ecosystem factor = utility of visual cues
\item mute aethestic = acceptable
\item mute aethestic = not acceptable
\item mute haptics = acceptable
\item mute haptics = not acceptable
\item mute form = acceptable
\item mute form = not acceptable
\item mute usability = acceptable
\item mute usability = not acceptable
\item mute concern = privacy protection
\item remote plug aethestic = acceptable
\item remote plug aethestic = not acceptable
\item remote plug form = acceptable
\item remote plug form = not acceptable
\item remote plug haptics = acceptable
\item remote plug haptics = not acceptable
\item remote plug usability = acceptable
\item remote plug usability = not acceptable
\item remote plug concern = boot up time
\item obfuscator aethestic = acceptable
\item obfuscator aethestic = not acceptable
\item obfuscator form = acceptable
\item obfuscator form = not acceptable
\item obfuscator haptics = acceptable
\item obfuscator haptics = not acceptable
\item obfuscator usability = acceptable
\item obfuscator usability = unsure
\item obfuscator usability = not acceptable
\item obfuscator concern = animals
\item obfuscator concern = harm device
\item ideal solution interface = voice
\item ideal solution interface = hands free
\item ideal solution interface = app
\item ideal solution integration = built in
\item ideal solution integration = bolt on
\item ideal solution form = minimal
\item ideal solution form = distributed for devices
\item ideal solution aethestic = multifunctional
\item ideal solution haptics = important
\item ideal solution haptics = not important
\item ideal solution ux = minimal interaction frequency
\item ideal solution ux = no downtime
\item ideal solution ux = no single point control
\item ideal solution ux = single point control
\item ideal solution other = all local
\item ideal solution developer = first party
\item ideal solution developer = third party
\item decision factor = cost
\item decision factor = privacy awareness
\end{enumerate}

\section{Items in the Commercial Market}

We provide screenshots of several cases/sleeves used to encase the Amazon Echo \va. Similar products can be found for the Google \va as well. 



\begin{figure}
 \centering
  \includegraphics[width=\columnwidth]{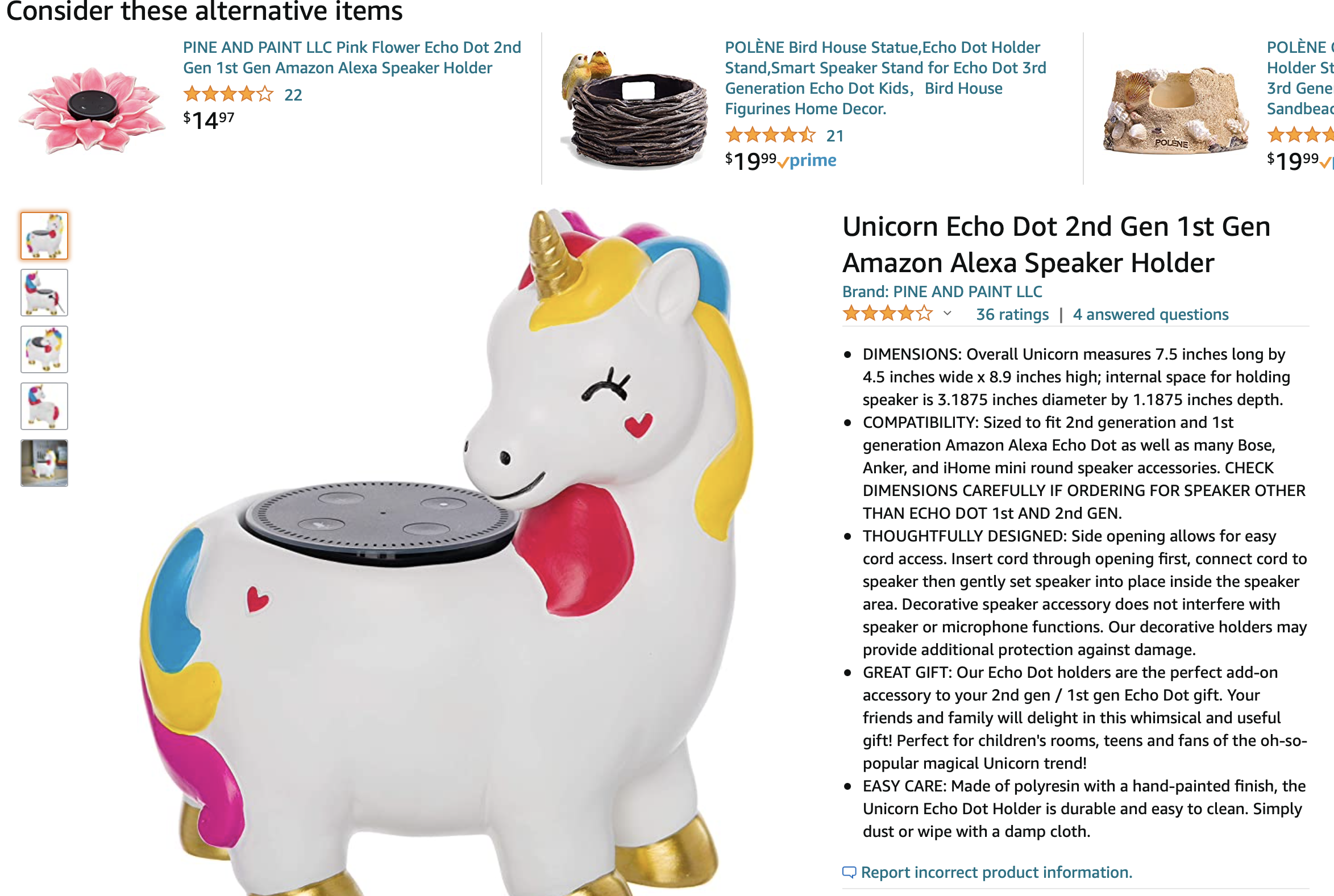}
  \caption{A case-like enclosing for Amazon Echo, on Amazon. \vspace*{-6mm}}
  \label{fig:amazon2}
\end{figure} 

\begin{figure}
 \centering
  \includegraphics[width=\columnwidth]{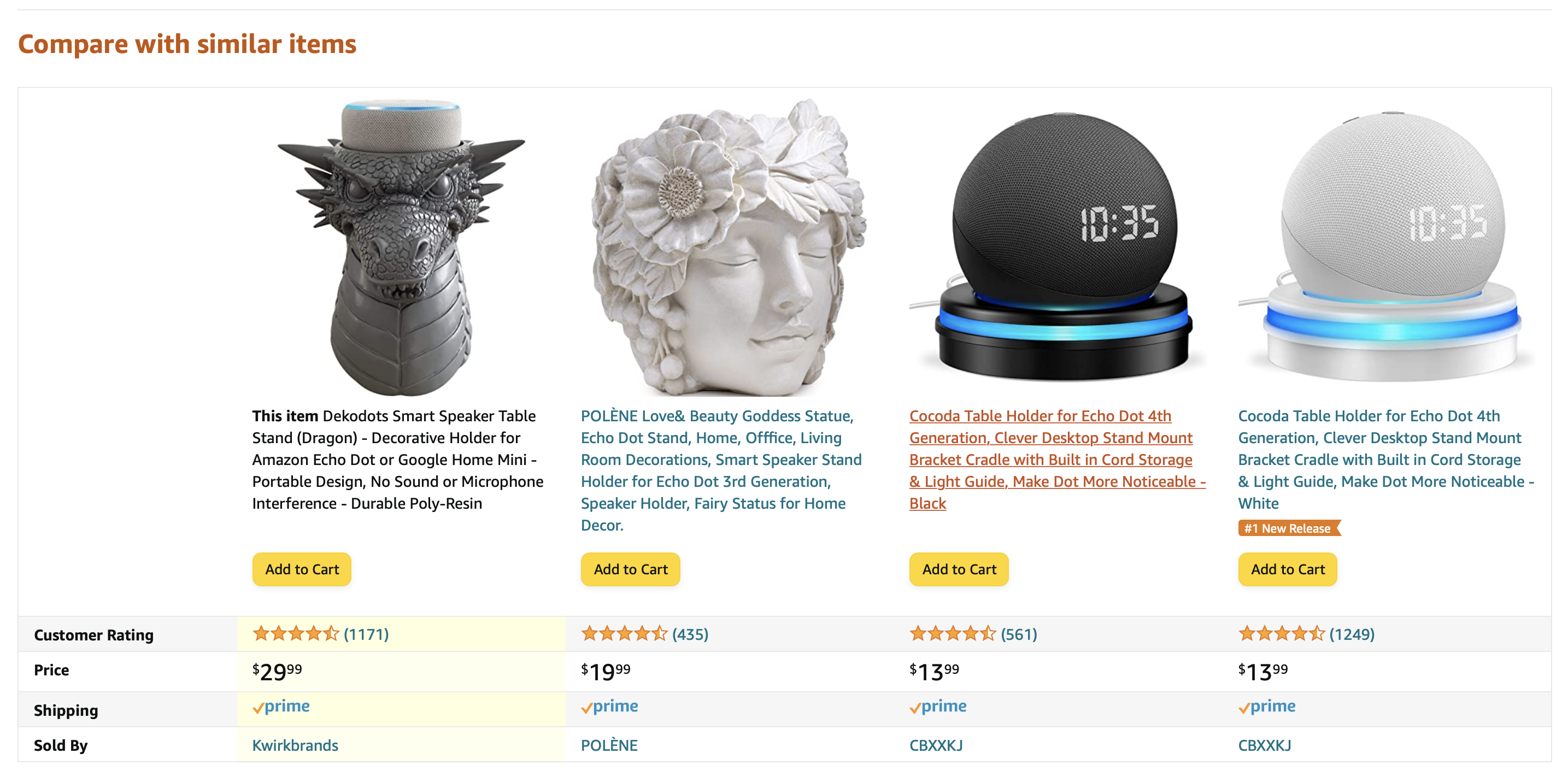}
  \caption{Case-like enclosing recommended by Amazon, for Amazon Echo, on Amazon.\vspace*{-6mm}}
  \label{fig:amazon1}
\end{figure}

\end{document}